\documentclass[aps,prb,reprint,groupedaddress]{revtex4-2}

\usepackage{graphicx}

\begin{document}

%\preprint{}


\title{Superheating fields of semi-infinite superconductors and layered superconductors in the diffusive limit: 
structural optimization based on the microscopic theory}


\author{Takayuki Kubo}
\email[]{kubotaka@post.kek.jp}
%\thanks{}
\affiliation{High Energy Accelerator Research Organization (KEK), Tsukuba, Ibaraki 305-0801, Japan}
\affiliation{The Graduate University for Advanced Studies (Sokendai), Hayama, Kanagawa 240-0193, Japan}


%\date{\today}

\begin{abstract}
We investigate the superheating fields $H_{sh}$ of semi-infinite superconductors and layered superconductors in the diffusive limit by using the well-established quasiclassical Green's function formalism of the BCS theory. 
The coupled Maxwell-Usadel equations are self-consistently solved to obtain the spatial distributions of the magnetic field, screening current density, penetration depth, and pair potential.
We find the superheating field of a semi-infinite superconductor in the diffusive limit is given by $H_{sh} = 0.795 H_{c0}$ at the temperature $T \to 0$. 
Here $H_{c0}$ is the thermodynamic critical-field at the zero temperature. 
Also, we evaluate $H_{sh}$ of layered superconductors in the diffusive limit as functions of the layer thicknesses ($d$) 
and identify the optimum thickness that maximizes $H_{sh}$ for various materials combinations. 
Qualitative interpretation of $H_{sh}(d)$ based on the London approximation is also discussed. 
The results of this work can be used to improve the performance of superconducting rf resonant cavities for particle accelerators. 
\end{abstract}

\maketitle


%%%%%%%%%%%%%%%%%%%
%%%%%%%%%%%%%%%%%%%
\section{Introduction}
%%%%%%%%%%%%%%%%%%%
%%%%%%%%%%%%%%%%%%%

The superconducting radio-frequency (SRF) resonant-cavity~\cite{2017_Padamsee, 2017_Gurevich_SUST} is the crucial component of modern particle accelerators, 
which efficiently imparts the electromagnetic energy to charged particles via the rf electric field. 
The accelerating gradient $E_{acc}$, namely, the average electric field that charged particles see during transit, 
is proportional to the amplitude $H_0$ of the rf magnetic field at the inner surface of the cavity, e.g., $\mu_0 H_0 = g E_{acc}$ and $g= 4.26\,{\rm mT/(MV/m)}$ for the TESLA-shape cavity. 
Today, the best Nb cavities can reach $\mu_0 H_0 \sim 200\,{\rm mT}$, which corresponds to $E_{acc} \simeq 40$-$50\,{\rm MV/m}$~\cite{2007_Geng, 2013_Singer, 2014_Kubo_IPAC, 2017_Grassellino, 2018_Dhakal}.

\begin{figure}[t]
   \begin{center}
   \includegraphics[width=0.95\linewidth]{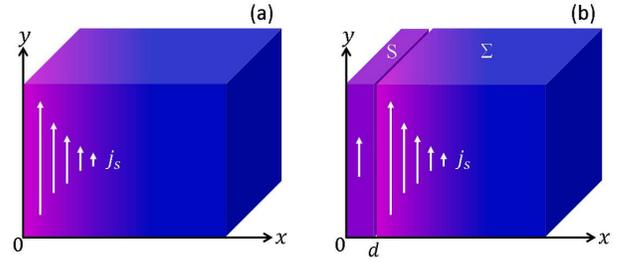}
\end{center}\vspace{0 cm}
   \caption{
(a) Semi-infinite superconductor occupying $x \ge 0$. 
(b) Layered structure that consists of a thin superconducting layer (S) and a superconducting substrate (${\rm \Sigma}$). 
Here, we assume S is completely decoupled from ${\rm \Sigma}$: 
we do not consider neither the proximity effect nor any electron transfer between S and ${\rm \Sigma}$. 
The magnetic field $H_0$ is applied parallel to $z$ axis, 
and the screening current flows parallel to $y$ axis.  
   }\label{fig1}
\end{figure}

The ultimate limit of $H_0$ is thought to be around $H_c$, 
irrespective of whether the cavity material is a type-I or a type-II superconductor. 
Here, $H_c$ is the thermodynamic critical field.  
This limitation comes from the fact that an SRF cavity is operated under the Meissner state, 
and the upper critical field $H_{c2}$ is irrelevant to SRF in contrast to some dc applications. 
Let us consider a semi-infinite superconductor in the Meissner state shown in Fig.~\ref{fig1} (a) and suppose the penetration depth is given by $\lambda$. 
The external magnetic field $H_0 \sim H_c$ induces the screening current density $j_s \sim H_c/\lambda$ at the surface, 
which is close to the depairing current density $j_d$, the stability limit of the superfluid flow. 
Hence, $H_0$ cannot substantially exceed $H_c$ as long as a simple semi-infinite superconductor is used. 
The value of $H_0$ which makes the Meissner state absolutely unstable is the so-called superheating field $H_{sh} \, (\sim H_c)$.

In the Ginzburg-Landau (GL) regime, $H_{sh}$ of a semi-infinite superconductor has been thoroughly investigated~\cite{1968_Kramer, 2011_Transtrum, 2016_Liarte}. 
However, the GL results are valid at a temperature $T$ close to the critical temperature $T_c$, 
while SRF cavities are operated at $T \ll T_c$ (e.g., $T/T_c \sim 0.1$-$0.2$ for Nb and ${\rm Nb_3 Sn}$ cavities). 
Microscopic calculations of $H_{sh}$, 
which are valid at an arbitrary temperature $0 < T < T_c$, 
have been carried out for extreme type-II superconductors, 
including clean-limit superconductors~\cite{1966_Galaiko, 2008_Catelani, 2017_Liarte_SUST}, 
superconductors including homogeneous~\cite{2012_Lin_Gurevich} and inhomogeneous impurities~\cite{2019_Sauls}, 
and dirty-limit superconductors with Dynes subgap states~\cite{2020_Kubo_Hsh}.

Besides the simple semi-infinite superconductor, 
the layered superconductor shown in Fig.~\ref{fig1} (b) has also attracted much attention from SRF researchers because of its potential for increasing the ultimate field-limit. 
Gurevich~\cite{2006_Gurevich} proposed the idea of multilayer coating, 
which introduces a higher-$H_c$ superconducting (S) layer formed on the top of the superconducting substrate (${\rm \Sigma}$). 
Here, the S layer is decoupled from ${\rm \Sigma}$ by an insulator layer or a natural oxide layer. 
Using the London theory, it was later shown~\cite{2014_Kubo} that, 
when the penetration depth of the S layer is larger than that of the substrate ${\rm \Sigma}$, 
a current counterflow induced by ${\rm \Sigma}$ leads to a suppression of the current density in the S layer, 
resulting in an enhancement of the ultimate field-limit. 
The enhancement is maximized when the S layer has the optimum thickness $d=d_m \sim \lambda^{\rm (S)}$~\cite{2014_Kubo}. 
It was also shown~\cite{2015_Posen} that the similar consequences result from the GL calculations, 
which are valid at $T\simeq T_c$. 
However, we should use the microscopic theory for quantitative analyses. 
Fortunately, for a clean-limit $s$-wave superconductor at $T=0$, 
analyses based on the microscopic theory are significantly simplified. 
The nonlinear Meissner effect is negligible in this regime, 
and the London equation is valid even under a strong current density close to $j_d$. 
Combining the current distribution obtained from the London equation and $j_d$ for a clean-limit superconductor calculated from the microscopic theory, 
the more quantitative theory of the optimum multilayer is obtained~\cite{2015_Gurevich}. 
These theoretical advances are discussed in detail in the review article~\cite{2017_Kubo_SUST}. 
Progress in experiments is summarized in Ref.~\cite{2017_Anne-Marie} (see also progress in the last several years, e.g., Refs.~\cite{2016_Tan, 2018_Junginger, 2019_Antoine, 2019_Kubo_JJAP, 2019_R_Ito_SRF, 2019_H_Ito_SRF, 2019_Katayama_SRF, 2019_Oseroff_SRF, 2019_Thoeng_SRF, 2019_Turner_SRF, 2019_Senevirathne_SRF, 2020_Ito}).

Also, Nb cavities processed by some materials-treatment methods (e.g., $120^{\,\circ}{\rm C}$ baking, nitrogen infusion, etc.) are known to have a thin dirty-layer (S) on the surface of the bulk Nb (${\rm \Sigma}$)~\cite{2006_Ciovati, 2014_Romanenko, 2017_Grassellino, 2018_Dhakal}, 
which can be modeled by the geometry shown in Fig.~\ref{fig1} (b). 
In fact, the calculations of the field-dependent nonlinear surface resistance~\cite{2017_Gurevich_Kubo, 2019_Kubo_Gurevich} have shown that layered structures can mitigate the quality factor degradation at high-fields. 
Moreover, it was shown~\cite{2017_Kubo_SUST, 2019_Sauls} that the thin dirty-layer at the surface improves $H_{sh}$ by the same mechanism as that in the S-${\rm \Sigma}$ heterostructure: 
a current counterflow induced by Nb substrate leads to a suppression of the current density in the dirty-Nb layer, 
resulting in an enhancement of $H_{sh}$. 
These theoretical results are qualitatively consistent with experiments~\cite{1998_Visentin, 1999_Lilje, 1999_Saito, 2017_Grassellino, 2018_Dhakal}. 
Other effects resulting from the materials treatment (e.g., effects on hydride precipitate~\cite{1976_Amano, 2012_Barkov}) may also play significant roles in the performance improvements, 
which can be incorporated considering imperfect surface-structures such as proximity-coupled normal layer on the surface~\cite{2017_Gurevich_Kubo, 2019_Kubo_Gurevich, 2020_Kubo_jd, 2020_Iavarone, 1996_Belzig, 1999_Belzig}.

Despite the extensive studies, the superheating fields of a simple semi-infinite superconductor and a layered heterostructure in the diffusive limit have not yet been studied. 
Also, the structural optimization of a layered heterostructure in the diffusive limit has not yet been done. 
In this regime, we can no longer use the London equation at $j_s \sim j_d$ due to the nonlinear Meissner effect~\cite{1995_Sauls, 2010_Groll, 2019_Sauls, 2020_Kubo_Hsh} in contrast to the clean-limit regime. 
We need to self-consistently solve the microscopic theory of superconductivity combined with the Maxwell equations, incorporating the current-induced pair-breaking effect and the resultant nonlinear Meissner effect~\cite{2012_Lin_Gurevich, 2019_Sauls, 2020_Kubo_Hsh}. 
In the present work, we evaluate $H_{sh}$ for these geometries and identify the optimum thicknesses of layered structures.

The paper is organized as follows. 
In Section II, we briefly review the Eilenberger-Usadel-Larkin-Ovchinnikov formalism of the BCS theory in the diffusive limit and express physical quantities with the Matsubara Green's functions. 
The solutions of the Usadel equation at $T=0$ and the analytical expression of the depairing current density are also summarized. 
In Sec. III, we consider a simple semi-infinite superconductor [see Fig.~\ref{fig1} (a)]. 
The coupled Maxwell-Usadel equations are self-consistently solved to obtain the spatial distributions of the magnetic field $H(x)$, the current density $j_s(x)$, pair potential $\Delta (x)$, and the penetration depth $\lambda(x)$. 
Then, the superheating field in the diffusive limit is derived. 
In Sec. IV, we consider a layered superconductor [see Fig.~\ref{fig1} (b)], self-consistently solve the coupled Maxwell-Usadel equations, and obtain the spatial distributions of $H(x)$, $j_s(x)$, $\Delta (x)$, and $\lambda(x)$. 
Then, we evaluate $H_{sh}$ as functions of the S-layer thickness $d$ for various material combinations and find the optimum thickness $d_m$. 
Qualitative interpretation of the results are also discussed using an approximate formula of $H_{sh}(d)$. 
In Sec. V, we discuss the implications of our results.

%%%%%%%%%%%%%%%%%%
%%%%%%%%%%%%%%%%%%
%%%%%%%%%%%%%%%%%%
\section{Theory} \label{sec_solutions}
%%%%%%%%%%%%%%%%%%
%%%%%%%%%%%%%%%%%%
%%%%%%%%%%%%%%%%%%

%%%%%%%%%%%%%%%%%%
\subsection{Eilenberger-Usadel-Larkin-Ovchinnikov formalism}
%%%%%%%%%%%%%%%%%%

Let us briefly summarize the well-established Eilenberger-Usadel-Larkin-Ovchinnikov formalism of the BCS theory in the diffusive limit~\cite{1999_Belzig, 1968_Eilenberger, 1969_LO, 1970_Usadel, Kopnin}. 
Here we assume the current distribution varies slowly over the coherence length. 
Then, the normal and anomalous quasiclassical Matsubara Green's functions $G_{\omega_n}=\cos\theta$ and $F_{\omega_n}=\sin\theta$ and the pair potential $\Delta$ obey
\begin{eqnarray}
&&\biggl( \Delta - \frac{s}{\sqrt{1+ \cot^2 \theta}} \biggr) \cot\theta = \hbar \omega_n , 
\label{thermodynamic_Usadel} \\
&&\ln \frac{T_{c}}{T} = 2 \pi k_B T \sum_{\omega_n >0} \biggl( \frac{1}{\hbar \omega_n} - \frac{\sin\theta}{\Delta} \biggr) .
\label{self-consistency}
\end{eqnarray}
Here 
$s =  (q/q_{\xi})^2 \Delta_0$ is the superfluid flow parameter, 
$\Delta_0 = \Delta(s, T)|_{s=0,\,T=0}$ is the BCS pair potential for the zero-current state at $T=0$, 
$\hbar q$ is the superfluid momentum, 
$q_{\xi} = \sqrt{2\Delta_0/\hbar D}$ is the inverse of the coherence length, 
$D$ is the electron diffusivity, 
$\hbar \omega_n = 2\pi k_B T(n + 1/2)$ is the Matsubara frequency, 
$k_B T_{c}= \Delta_0 \exp(\gamma_E)/\pi  \simeq \Delta_0/1.76$ is the BCS critical temperature, 
and $\gamma_E=0.577$ is the Euler constant. 
The penetration depth $\lambda$ and the magnitude of supercurrent density $j_s$ are given by
\begin{eqnarray}
&&%\frac{n_s(s, T)}{n_{s0}} = \frac{\lambda_0^2}{\lambda^2(s, T)}
\frac{\lambda_0^2}{\lambda^2(s,T)} = \frac{4k_B T}{\Delta_0} \sum_{\omega_n > 0} \sin ^2 \theta 
, \label{superfluid_density} \\
&&\frac{j_s (s, T)}{j_{s0}} 
= \sqrt{\frac{\pi s}{\Delta_0}} \frac{\lambda_0^2}{\lambda^2(s,T)} . \label{supercurrent} 
\end{eqnarray}
Here $\lambda_0 = \lambda(0,0) = \sqrt{\hbar/\pi \mu_0 \Delta_0 \sigma_n}$ is the BCS penetration depth at $T=0$, 
$\sigma_n = 2N_0 D e^2$ is the normal state conductivity, 
$N_0$ is the normal state density of states at the Fermi energy, 
$j_{s0} = H_{c0}/\lambda_0 = \sqrt{\pi} |e| N_0 D \Delta_0 q_{\xi}$, and
$H_{c0}=\sqrt{N_0/\mu_0} \Delta_0$ is the BCS thermodynamic critical field at $T=0$.

In the geometries shown in Figs.~\ref{fig1} (a) and \ref{fig1} (b), 
the magnetic field and the superfluid flow depend on the depth $x$ from the surface. 
These $x$ dependences are determined from the Maxwell equations, 
$j_s = -\partial_x H$ and $\mu_0 H=(\hbar/2|e|) \partial_x q$, namely,
\begin{eqnarray}
&&\frac{\partial^2 q}{\partial x^2} = \frac{q}{\lambda^2 (s, T)} , \label{Usadel_London} \\
&& \frac{H}{H_{c0}} = \sqrt{\pi}  \frac{\partial (q/q_{\xi})}{\partial (x/\lambda_0)} . \label{H}
\end{eqnarray}
Suppose the magnetic field at the surface is given by $H_0$.  
Then, the boundary conditions can be written as
\begin{eqnarray}
H(0) =H_0, \hspace{1.8cm} \lim_{x\to \infty} q(x) \to 0.   \label{boundary_conditions_1} 
\end{eqnarray}
For the geometry shown in Fig.~\ref{fig1} (b), we have the additional boundary conditions at the S-${\rm \Sigma}$ interface, 
\begin{eqnarray}
H(d_{-}) = H(d_{+}), \hspace{1cm} q(d_{-}) = q(d_{+}) .\label{boundary_conditions_2}
\end{eqnarray}
Here $d_{\pm} = d \pm 0$: 
we assume the thickness of the insulator or natural oxide layer separating S and ${\rm \Sigma}$ is negligible compared with $\lambda$ but thick enough to electrically decouple S and ${\rm \Sigma}$.

%%%%%%%%%%%%%%%%%%
\subsection{Solutions and depairing current at $T \to 0$}\label{sec_current_carrying}
%%%%%%%%%%%%%%%%%%

\begin{figure}[t]
   \begin{center}
   \includegraphics[width=0.48\linewidth]{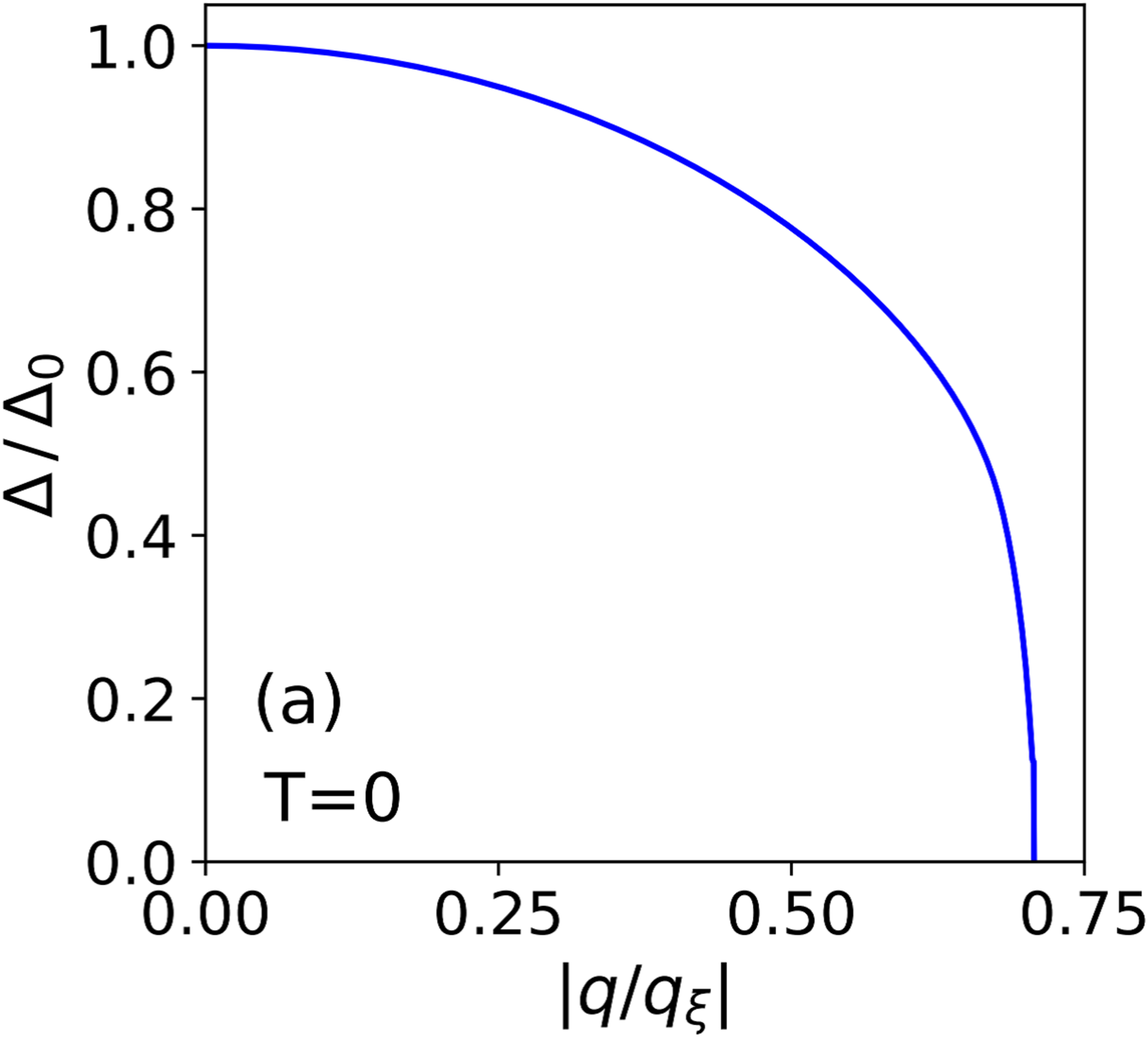}
   \includegraphics[width=0.48\linewidth]{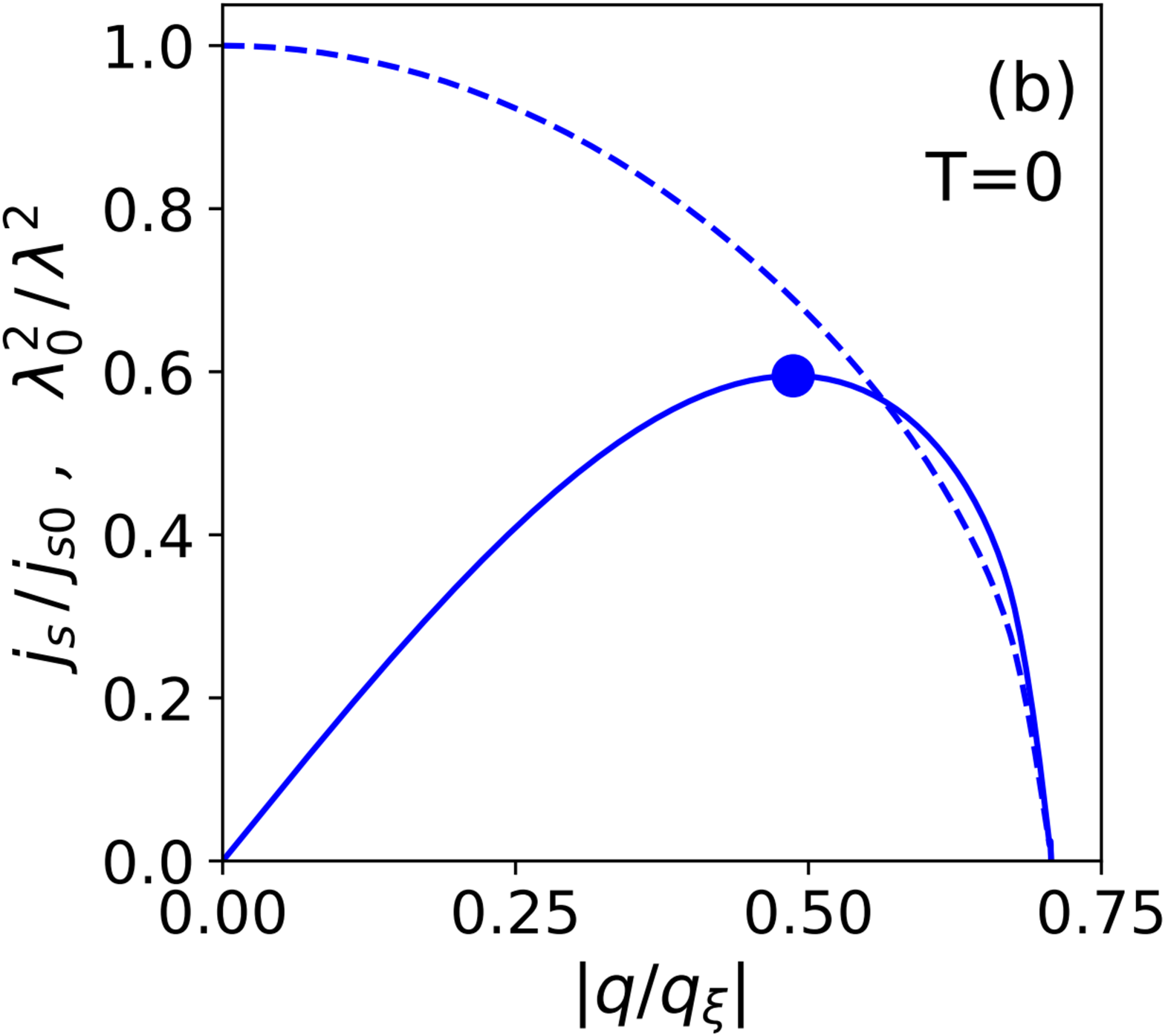}
\end{center}\vspace{0 cm}
   \caption{
(a) Pair potential $\Delta$, (b) superfluid density $\lambda_0^2/\lambda^2$ and supercurrent density $j_s$ as a function of the superfluid momentum $|q/q_{\xi}| \, (= \sqrt{s/\Delta_0})$ calculated from the Usadel equation at $T\to 0$ [Eqs.~(\ref{self-consistency_Tzero_1}) and (\ref{superfluid_density_Tzero_1}), see Appendix~\ref{a1}]. 
For $0 \le |q/q_{\xi}| \le 0.675$, we can use Eqs.~(\ref{self-consistency_Tzero_2})-(\ref{js_Tzero_2}). 
The blob at $(0.487, 0.595)$ indicates the depairing momentum and depairing current density. 
   }\label{fig2}
\end{figure}

For $T \to 0$, the Matsubara sum is replaced with an integral, 
and Eqs.~(\ref{thermodynamic_Usadel})-(\ref{supercurrent}) reduce to the well-known formulas obtained by Maki many years ago~\cite{1963_Maki_I, 1963_Maki_II, 2012_Clem_Kogan, 2020_Kubo_Hsh} (see also Appendix~\ref{a1}),
\begin{eqnarray}
&& \frac{\Delta (s, 0)}{\Delta_0} = \exp\biggl[ -\frac{\pi s}{4\Delta (s, 0)}  \biggr], \label{self-consistency_Tzero_2} \\
&& \frac{\lambda_0^2}{\lambda^2(s, 0)}
= \frac{\Delta (s, 0)}{\Delta_0} \biggl[ 1 -\frac{4s}{3\pi \Delta(s,0)} \biggr]  , \label{superfluid_density_Tzero_2} \\
&& j_s (s,0) = \sqrt{\frac{\pi s}{\Delta_0}} \frac{\Delta (s, 0)}{\Delta_0} \biggl[ 1 -\frac{4s}{3\pi \Delta(s,0)} \biggr] \frac{H_{c0}}{\lambda_0},   \label{js_Tzero_2}
\end{eqnarray}
for $0 \le s \le \Delta(s,0)$, namely, $0 \le s \le 0.456\Delta_0$ or $0 \le |q/q_{\xi}| \le 0.675$. 
Shown in Fig.~\ref{fig2} are $\Delta$, $\lambda$, and $j_s$ at $T=0$ as functions of $|q|$. 
While $\Delta$ and the superfluid density $\lambda_0^2/\lambda^{2}$ are monotonically decreasing functions, 
$j_s$ exhibits a non-monotonic behavior. 
At smaller $|q|$ regions, $j_s$ is proportional to $|q|$. 
As $|q|$ increases, $j_s$ becomes dominated by a rapid reduction of the superfluid density and ceases to increase. 
At a threshold value $q_d$, $j_s$ reaches the maximum value:  
the depairing current density $j_d$ [see the blob in Fig.~\ref{fig2} (b)].

Using Eqs.~(\ref{self-consistency_Tzero_2})-(\ref{js_Tzero_2}) and the condition $\partial_s j_s =0$, 
we have~\cite{1963_Maki_I, 1963_Maki_II, 1980_Kupriyanov} 
\begin{eqnarray}
&&j_d(0) = \sqrt{\frac{\pi s_d}{\Delta_0}} \frac{\Delta_d}{\Delta_0} \biggl( 1 -\frac{4\zeta_{d}}{3\pi} \biggr) \frac{H_{c0}}{\lambda_0} =0.595 \frac{H_{c0}}{\lambda_0} , \label{jd_approx} \\
&&\zeta_{d}=\frac{s_d}{\Delta_d}= \frac{2}{\pi} + \frac{3\pi}{8} - \sqrt{\biggl( \frac{2}{\pi} + \frac{3\pi}{8} \biggr)^2 -1}  = 0.300 ,  
\\
&& \Delta_{d} = \Delta_0 e^{-\frac{\pi}{4} \zeta_{d}} = 0.790 \Delta_0, \label{Delta_d0} \\
&& s_{d} = \Delta_{d} \zeta_{d} = 0.237 \Delta_0, \label{s_d0} \\
&& q_d / q_{\xi} = \sqrt{s_d/\Delta_0} = 0.487 ,
\end{eqnarray}
which are the well-known formula of the depairing current density for a dirty BCS superconductor (see also Refs.~\cite{2012_Clem_Kogan, 2020_Kubo_Hsh}).

In the following, we use $\Delta_{0}$ as a unit of energy and use dimensionless quantities 
$\tilde{s} = s/\Delta_{0}$, 
$\tilde{\omega}_n= \hbar \omega_n/\Delta_{0}$, 
$\tilde{\Delta}= \Delta/\Delta_{0}$, 
$\tilde{T}= k_B T/\Delta_{0}$, etc. 
For brevity, we omit all these tildes. 
Also, since we are interested in $H_{sh}$ at the operating temperature of SRF cavities ($T \ll T_c$), 
we consider $T \to 0$ for simplicity.

%%%%%%%%%%%%%%%%%%
%%%%%%%%%%%%%%%%%%
%%%%%%%%%%%%%%%%%%
\section{Semi-infinite superconductor} \label{sec_bulk}
%%%%%%%%%%%%%%%%%%
%%%%%%%%%%%%%%%%%%
%%%%%%%%%%%%%%%%%%

%%%%%%%%%%%%%%%%%%
\subsection{Spatial distributions of $H$, $j_s$, $\Delta$, and $\lambda$} \label{Sec_bulk_distribution}
%%%%%%%%%%%%%%%%%%

\begin{figure}[t]
   \begin{center}
   \includegraphics[width=0.48\linewidth]{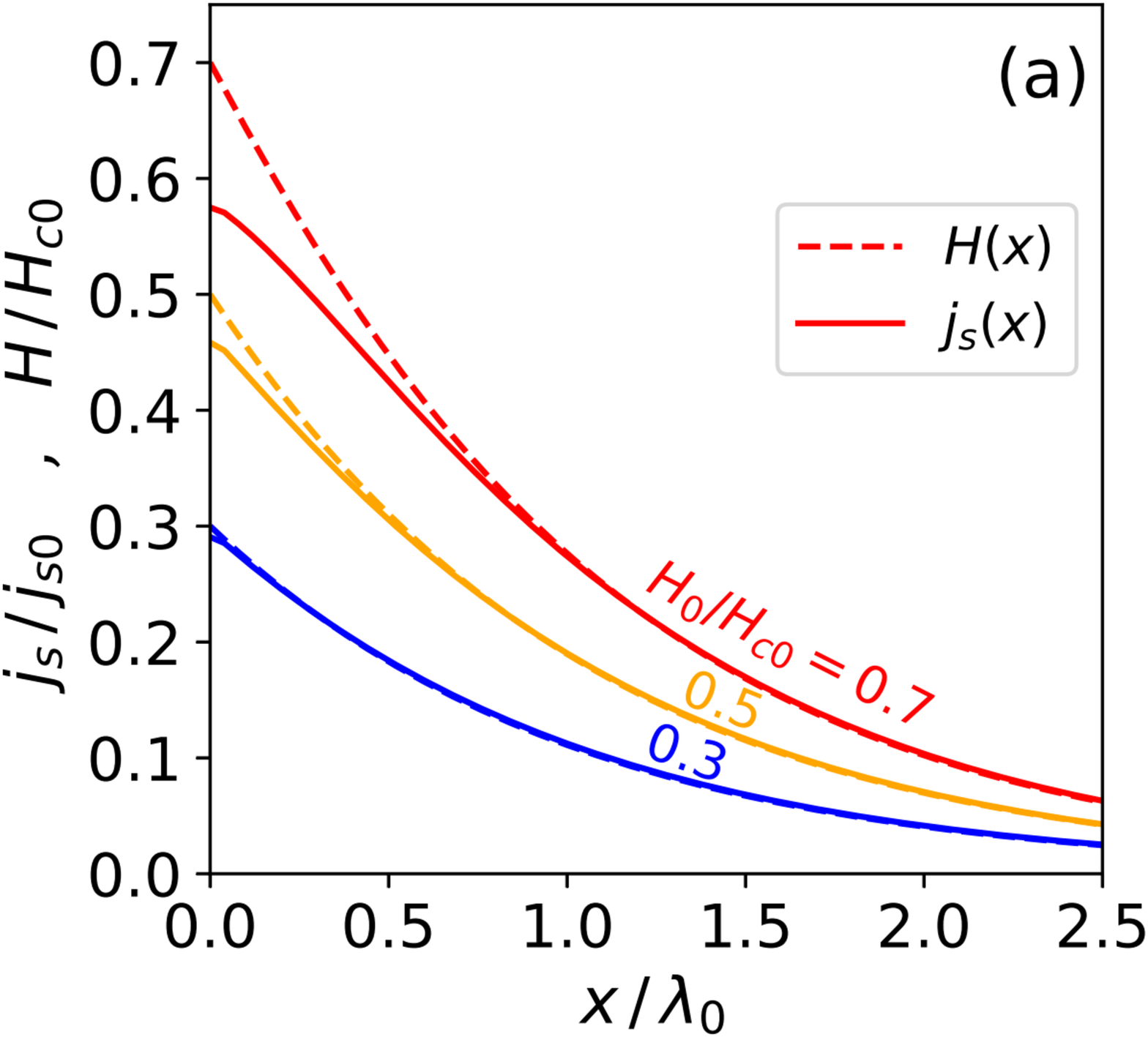}
   \includegraphics[width=0.48\linewidth]{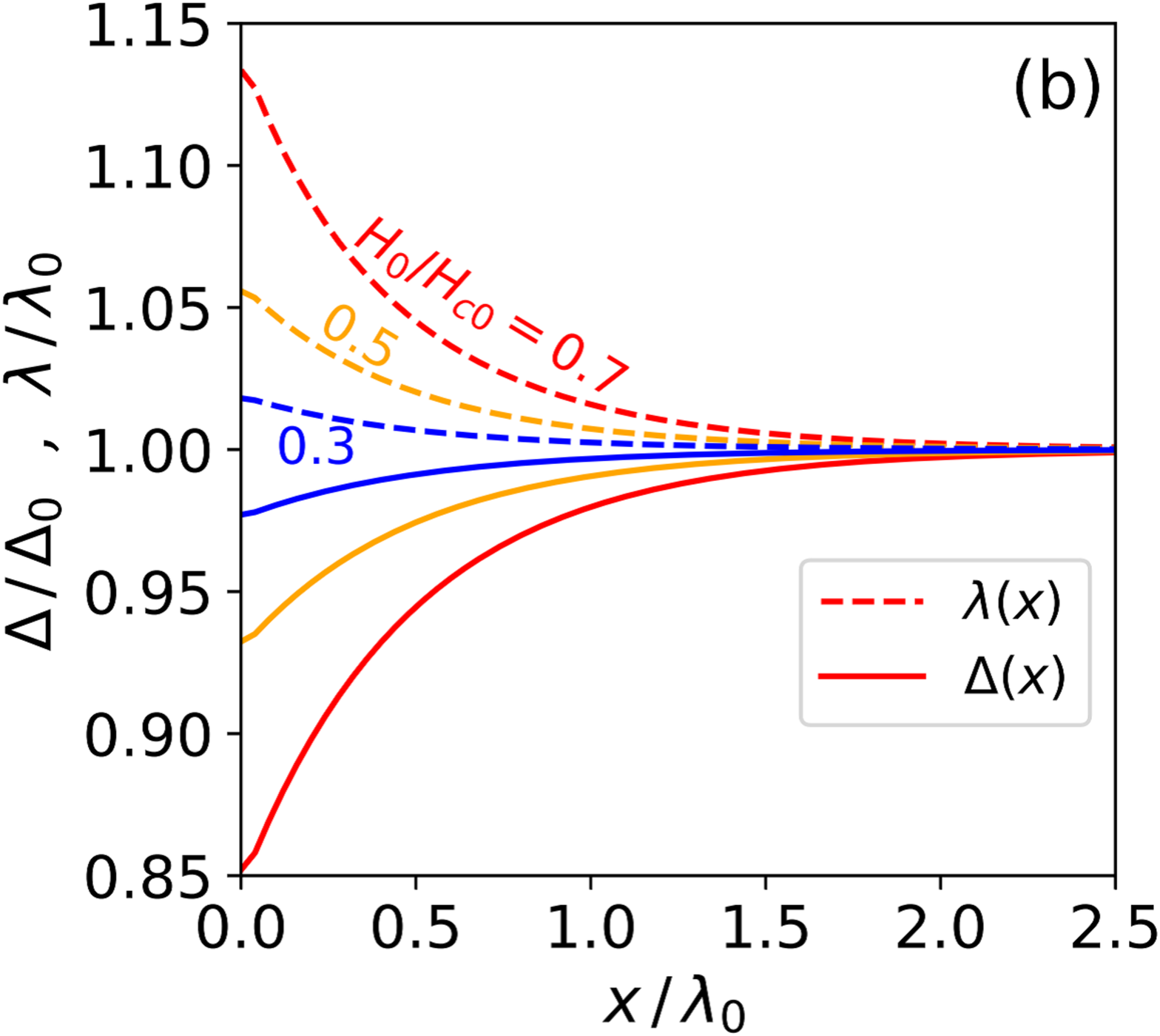}
\end{center}\vspace{0 cm}
   \caption{
Spatial distributions of (a) $H$, $j_s$, (b) $\lambda$, and $\Delta$ obtained from the self-consistent solutions of the coupled Maxwell-Usadel equations [Eqs.~(\ref{Usadel_London})-(\ref{js_Tzero_2})]. 
For $H_0 \sim H_{c0}$ (red), the nonlinear Meissner effect manifests itself in the vicinity of the surface. 
   }\label{fig3}
\end{figure}

First, consider the semi-infinite superconductor shown in Fig.~\ref{fig1} (a) and self-consistently solve the coupled Maxwell-Usadel equations at $T \to 0$ [Eqs.~(\ref{Usadel_London})-(\ref{js_Tzero_2})]. 
Shown in Fig.~\ref{fig3} (a) are the distributions of $H(x)$ and $j_s(x)$. 
For $H_0 \ll H_{c0}$ (blue), 
we have the solid and dashed curves that almost overlap, 
which can be understood as follows. 
Since the current density is so small that the nonlinear Meissner effect is negligible, 
the London theory is applicable. 
Solving the London equation, we obtain $H(x)/H_{c0}=j_s(x)/j_{s0} = \exp (-x/\lambda_0)$, 
consistent with the numerical solutions. 
As $H_0$ increases (orange and red), however, 
the nonlinear Meissner effect manifests itself, and the London theory is not applicable. 
In fact, the solid and dashed curves no longer overlap. 
Shown in Fig.~\ref{fig3} (b) are the penetration depth $\lambda(x)$ and the pair potential $\Delta(x)$. 
The pair potential (penetration depth) is decreased (increased) at the surface due to the strong pair-breaking current and recovers at deeper regions where the current density is exponentially small.

%%%%%%%%%%%%%%%%%%
\subsection{Superheating field}\label{sec_bulk_Hsh}
%%%%%%%%%%%%%%%%%%

For a simple semi-infinite superconductor in the diffusive limit, 
in which the current density is a monotonically decreasing function of $x$ (see Fig.~\ref{fig3}), 
the superheating field $H_{sh}$ is given by $H_0$ which induces $j_s(x)|_{x=0} = j_d$. 
Then, we can derive a simple formula of $H_{sh}$. 
Integrating both the sides of Eq.~(\ref{Usadel_London}) from $x=0$ to $\infty$, 
we obtain $q'(0)^2 = -2 \int_0^{\infty} q q' \lambda^{-2}(s,T) dx$. 
Using Eqs.~(\ref{H}) and (\ref{boundary_conditions_1}), 
we find the relation between $H_0$ and $s(x)|_{x=0}$: 
$(H_0/H_{c0})^2=\pi \int_0^{s(0)} ds [\lambda_0/\lambda(s,T)]^2$. 
Substituting the depairing value $s_d$ into $s(0)$, we obtain the formula~\cite{2020_Kubo_Hsh} 
\begin{eqnarray}
H_{sh} (T)
= H_{c0} \sqrt{ \pi  \int_0^{s_d} \!\!\!\!ds  \frac{\lambda_0^2}{\lambda^2(s,T)}   } ,
\label{Hsh_formula}
\end{eqnarray}
which is valid for an arbitrary $T$. 
Note that Eq.~(\ref{Hsh_formula}) reproduces the GL superheating field $H_{sh}(T)=(\sqrt{5}/3) H_c(T)$ at $T\simeq T_c$~\cite{2020_Kubo_Hsh}, 
consistent with the previous studies~\cite{1968_Kramer, 2011_Transtrum}.

For $T=0$, 
substituting Eqs.~(\ref{superfluid_density_Tzero_2}) and (\ref{s_d0}) into Eq.~(\ref{Hsh_formula}), 
we obtain $H_{sh}$ in the diffusive limit~\cite{2020_Kubo_Hsh}  
\begin{eqnarray}
H_{sh}(0) 
&=& H_{c0} \sqrt{ 1 - \biggl( 1 -\frac{\pi \zeta_{d}}{2} \biggr) e^{-\frac{\pi\zeta_{d}}{2}} -\frac{2}{3}s_{d}^2 } \nonumber \\
&=& 0.795 H_{c0} . \label{Hsh_zero}
\end{eqnarray}
This is slightly smaller than that of an extreme type-II ($\lambda/ \xi \gg 1$) superconductor in the clean limit~\cite{1966_Galaiko, 2008_Catelani}, 
\begin{eqnarray}
H_{sh}^{\rm clean} (0) 
= 0.84 H_{c0} ,  \label{Hsh_clean}
\end{eqnarray}
and consistent with the previous study on the effect of nonmagnetic impurities~\cite{2012_Lin_Gurevich}, 
in which $H_{sh} \,(\sim 0.8 H_{c0})$ as a function of the mean free path $\ell$ takes its maximum value at $\ell =5.32 \xi_0= \ell_*$ and decreases with $\ell$ for $\ell < \ell_*$.

%%%%%%%%%%%%%%%%%%
%%%%%%%%%%%%%%%%%%
\section{Layered superconductors} \label{sec_layered}
%%%%%%%%%%%%%%%%%%
%%%%%%%%%%%%%%%%%%

Now we consider the layered heterostructure shown in Fig.~\ref{fig1} (b). 
The model parameters are summarized in Table.~1: 
the S-layer thickness $d$ and the three ratios of materials parameters, 
\begin{eqnarray}
r_{\Delta} = \frac{\Delta_0^{\rm (S)}}{\Delta_0^{\rm (\Sigma)}} , \hspace{0.7cm}
r_{H} = \frac{H_{c0}^{\rm (S)}}{H_{c0}^{\rm (\Sigma)}} , \hspace{0.7cm}
r_{\sigma} = \frac{\sigma_n^{\rm (S)}}{\sigma_n^{\rm (\Sigma)}} . \label{parameters}
\end{eqnarray}
Here $\Delta_0^{(i)}$ ($i={\rm S}, \, {\rm \Sigma}$) is the pair-potential in the zero-current state at $T=0$, 
$H_{c0}^{(i)}$ is the thermodynamic critical field at $T=0$, 
and $\sigma_n^{(i)}$ is the normal-state conductivity. 
The other materials parameters can be expressed using these parameters;  
e .g., $D^{\rm (S)}/D^{\rm (\Sigma)}=r_{\sigma} r_{\Delta}^2/r_H^2$, 
$\lambda_0^{({\rm S})}/\lambda_0^{({\rm \Sigma})} = 1/\sqrt{r_{\Delta} r_{\sigma}}$, etc.

\begin{table}[t]
\caption{\label{tab:table}
Parameters of the layered structure. 
}
\begin{tabular}{cc}
\hline
\hline
S layer thickness    & $d$, \\
Pair-potential ratio & $r_{\Delta} = \Delta_0^{\rm (S)}/\Delta_0^{\rm (\Sigma)}$,  \\
Critical-field ratio   & $r_{H} = H_{c0}^{\rm (S)}/H_{c0}^{\rm (\Sigma)}$, \\
Normal-conductivity ratio   & $r_{\sigma} = \sigma_n^{\rm (S)}/\sigma_n^{\rm (\Sigma)}$. \\
\hline
\hline
\end{tabular}
\end{table}

%%%%%%%%%%%%%%%%%%
\subsection{Spatial distributions of $H$, $j_s$, $\Delta$, and $\lambda$ in a layered superconductor} \label{sec_layered_distribution}
%%%%%%%%%%%%%%%%%%

\begin{figure}[t]
   \begin{center}
   \includegraphics[width=0.48\linewidth]{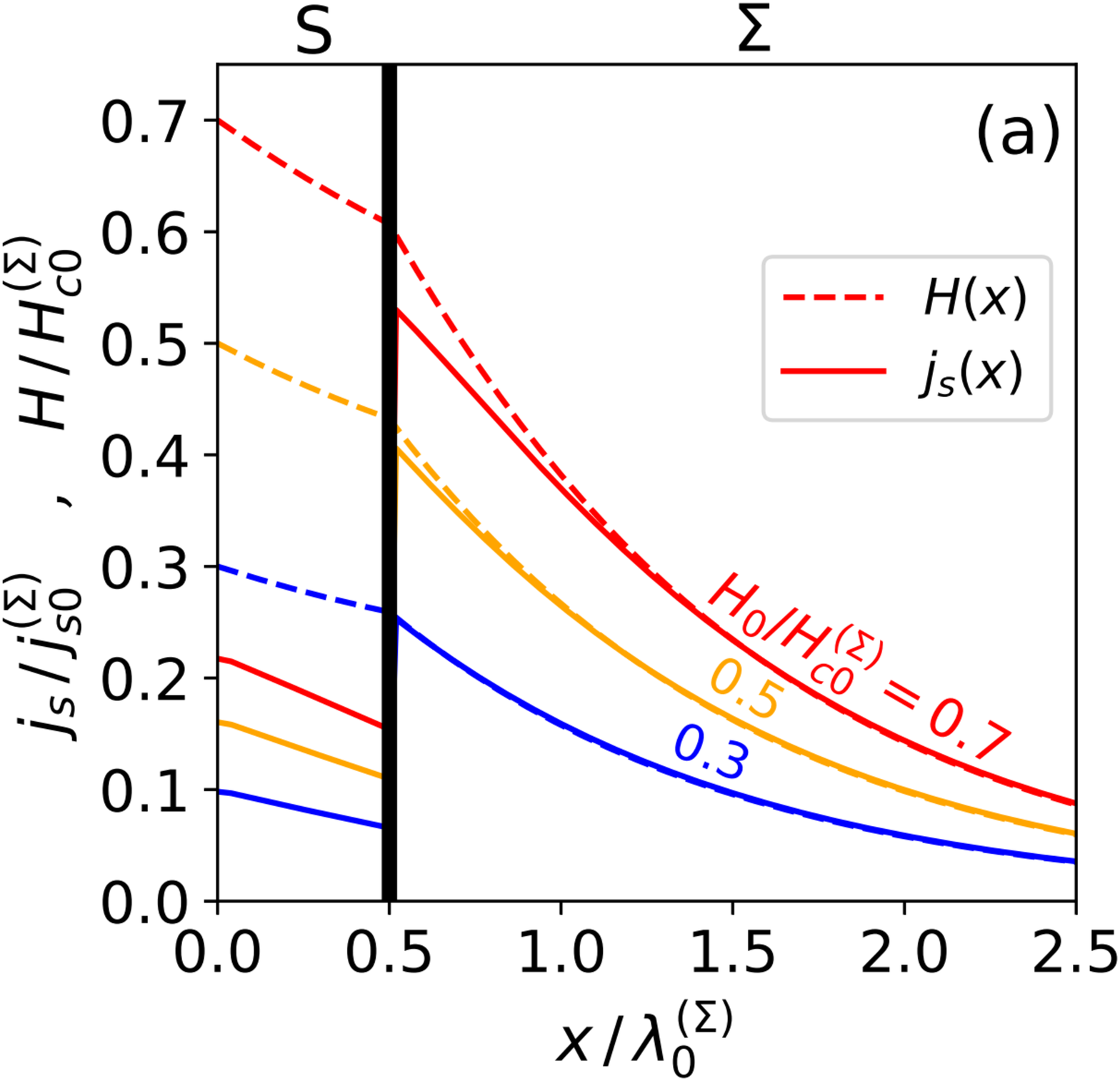}
   \includegraphics[width=0.48\linewidth]{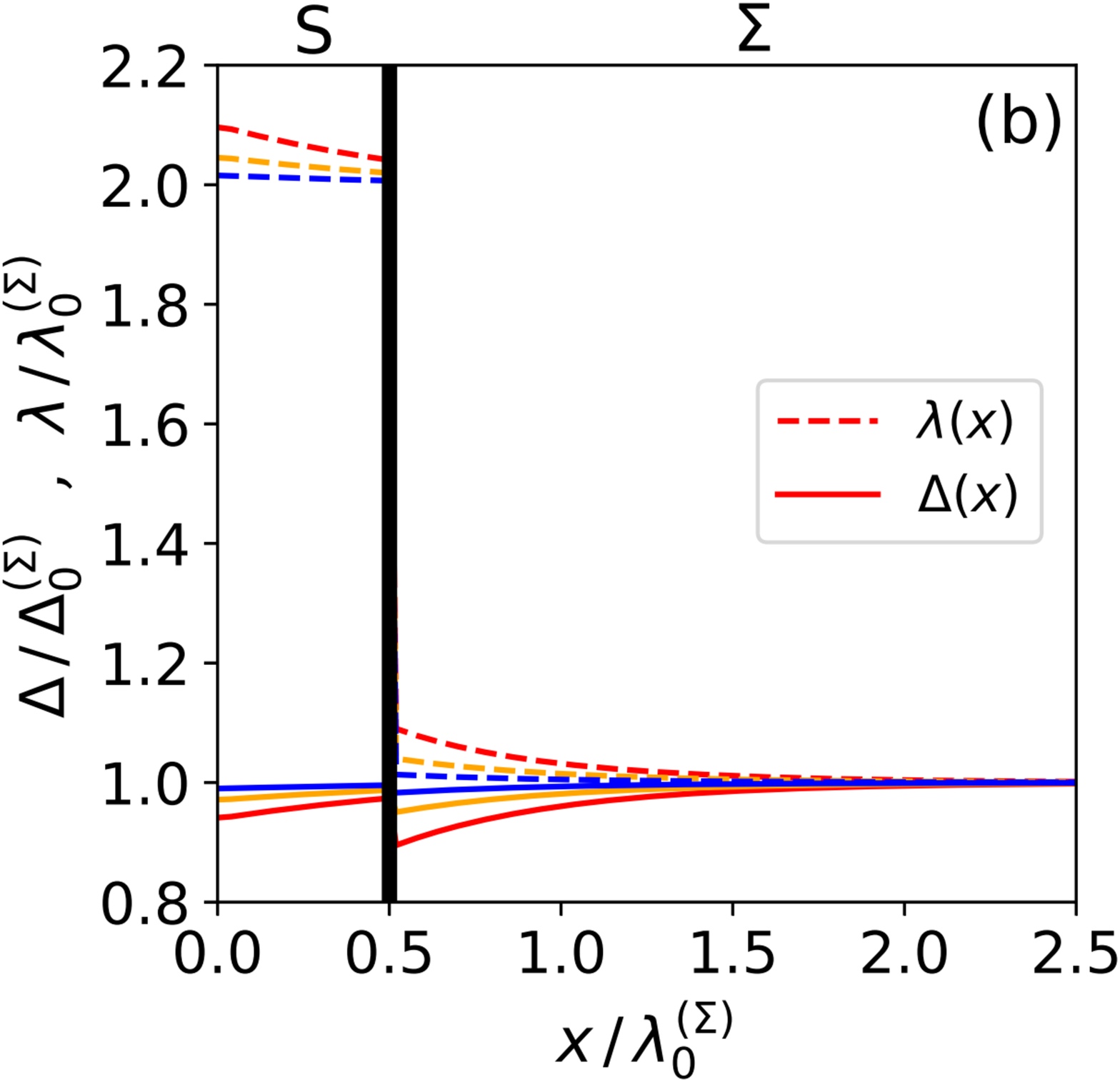}
\end{center}\vspace{0 cm}
   \caption{
Spatial distributions of (a) $H$, $j_s$, (b) $\Delta$, and $\lambda$ in a layered superconductor calculated for $d=0.5 \lambda_0^{\rm (\Sigma)}$, $r_{\Delta}=r_{H}=1$, and $r_{\sigma} =0.25$. 
The penetration depth of the S layer in the zero-current state is given by $\lambda_0^{\rm (S)}=\lambda_0^{\rm (\Sigma)}/\sqrt{r_{\Delta}r_{\sigma}}=2\lambda_0^{\rm (\Sigma)}$. 
   }\label{fig4}
\end{figure}

We can consider any materials combination, but here we focus on the simplest example that captures the striking feature of the layered structure: the surface-current-reduction effect. 
Let us assume that the S layer is made of the same material as the ${\rm \Sigma}$ region but has a different concentration of nonmagnetic impurities, 
e.g., ${\rm \Sigma}$ and S are a bulk Nb and a dirtier Nb-layer, respectively.
We can fully solve the coupled Maxwell-Usadel equations at $T \to 0$ [Eqs.~(\ref{Usadel_London})-(\ref{js_Tzero_2})] in  the similar way as done for a semi-infinite superconductor in Sec.~\ref{Sec_bulk_distribution}. 
Shown in Fig.~\ref{fig4} are the distributions of $H(x)$, $j_s(x)$, $\lambda(x)$, and $\Delta(x)$ calculated for $d=0.5 \lambda_0^{\rm (\Sigma)}$, $r_{\Delta}=r_{H}=1$, and $r_{\sigma} =0.25$. 
In the S layer, the magnetic field $H$ (dashed curves) slowly attenuates as $x$ increases, 
and the current density $j_s=-\partial_x H$ (solid curves) is significantly suppressed. 
As a result, $\Delta$ in the S region is less suppressed as compared with that in the ${\rm \Sigma}$ region, 
even though it is the S layer which is directly exposed to the external magnetic field [see the solid curves in Fig.~\ref{fig4} (b)]. 
In the ${\rm \Sigma}$ region, 
$H(x)$, $j_s(x)$, $\lambda(x)$, and $\Delta(x)$ monotonically decay as $x$ increases.

The non-monotonic decay of the current density is a common feature in the S-${\rm \Sigma}$ hetero-structure in which S has a different penetration depth from ${\rm \Sigma}$~\cite{2014_Kubo, 2015_Posen, 2015_Gurevich, 2017_Kubo_SUST, 2019_Sauls}. 
When ${\rm \Sigma}$ has a shorter (longer) penetration depth than S, 
the magnetic field in the S layer decays slower (more rapid) than exponential, 
and the current density is suppressed (enhanced). 
Such reduction (enhancement) of the surface current results from a counterflow induced by the substrate ${\rm \Sigma}$. 
When the penetration depths in S and ${\rm \Sigma}$ are balanced, 
the magnetic field and the current density in the S layer exhibit the well-known exponential decay.

\begin{figure}[t]
   \begin{center}
   \includegraphics[width=0.48\linewidth]{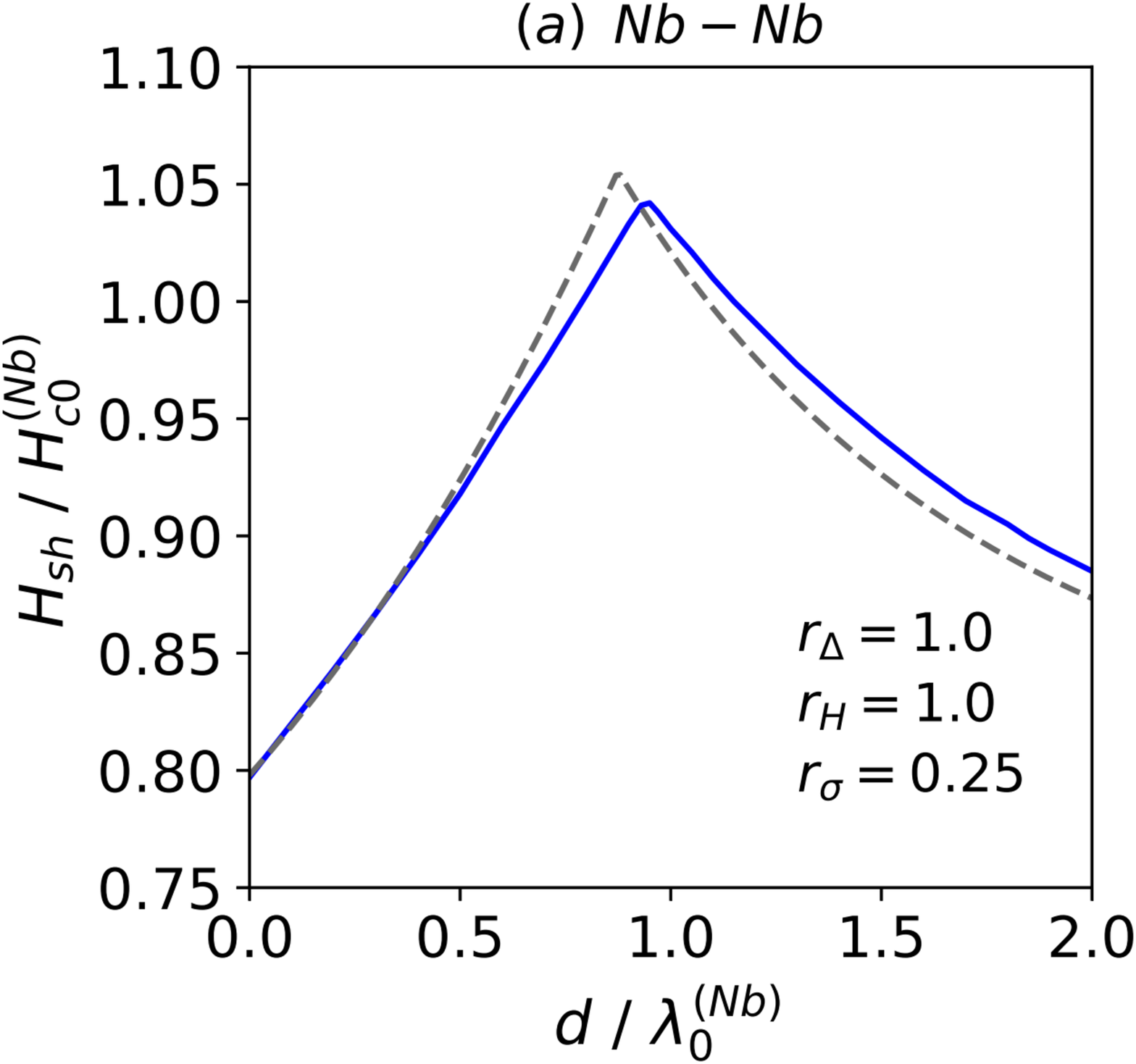}
   \includegraphics[width=0.48\linewidth]{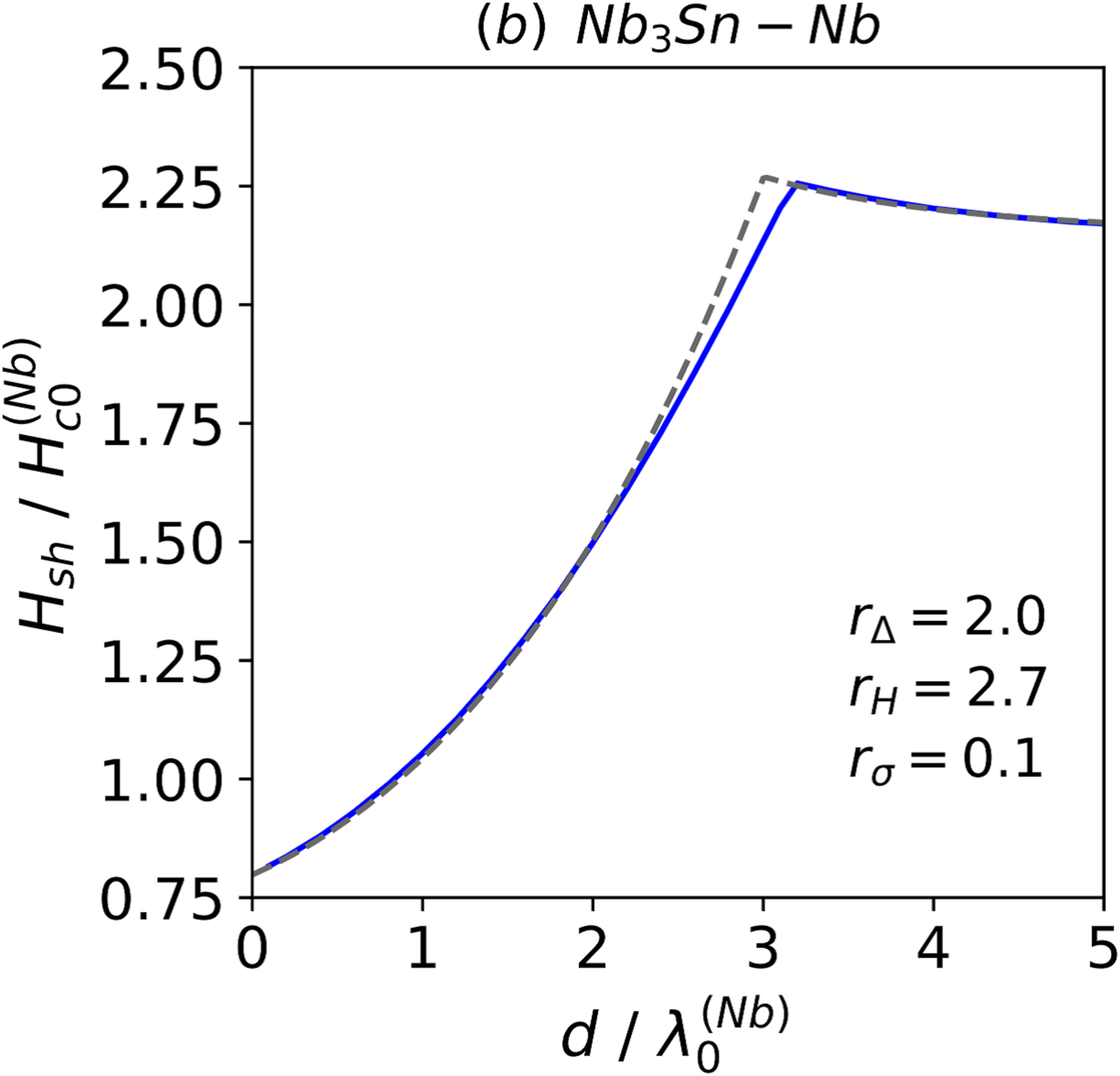}
   \includegraphics[width=0.48\linewidth]{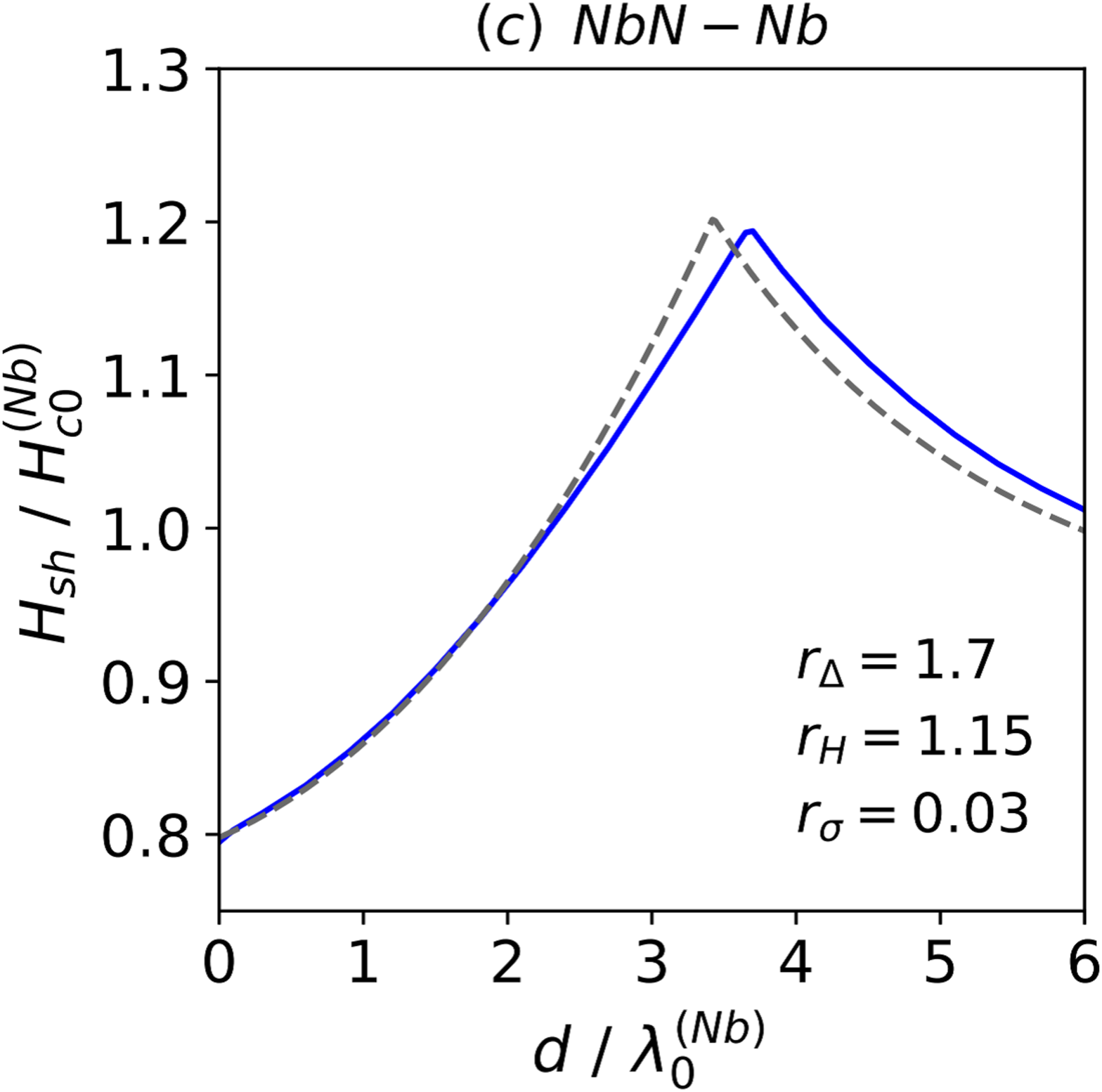}
   \includegraphics[width=0.48\linewidth]{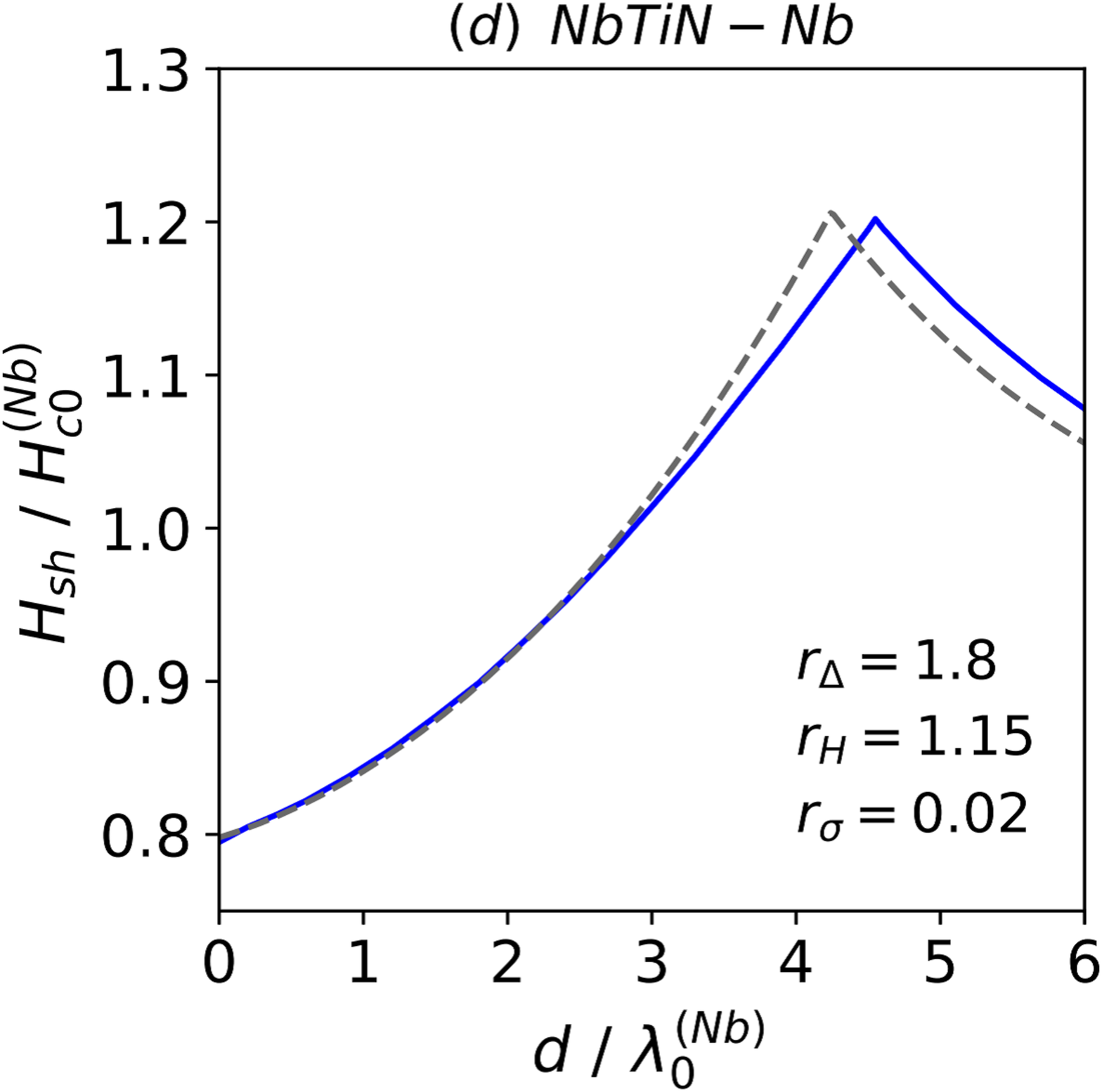}
\end{center}\vspace{0 cm}
   \caption{
Superheating field of layered structures as functions of $d$. 
The solid curves are obtained from the self-consistent solutions of the coupled Maxwell-Usadel equations at $T \to 0$ [Eqs.~(\ref{Usadel_London})-(\ref{js_Tzero_2})]. 
The dashed curves are calculated from the approximate formula [Eq.~(\ref{Hsh_London})]. 
(a) Nb-Nb structure modeled by the parameter set $(r_{\Delta}, r_{H}, r_{\sigma})=(1,1,0.25)$. 
(b) ${\rm Nb_3 Sn}$-Nb structure modeled by $(r_{\Delta}, r_{H}, r_{\sigma})=(2,2.7,0.1)$. 
(c) NbN-Nb structure modeled by $(r_{\Delta}, r_{H}, r_{\sigma})=(1.7, 1.15, 0.03)$. 
(d) NbTiN-Nb structure modeled by $(r_{\Delta}, r_{H}, r_{\sigma})=(1.8, 1.15, 0.02)$. 
In these cases, the penetration depths of the S layer in the zero-current state are given by (a) $\lambda_0^{\rm (S)}=\lambda_0^{\rm (\Sigma)}/\sqrt{r_{\Delta}r_{\sigma}}=2\lambda_0^{\rm (\Sigma)}$, 
(b) $\lambda_0^{\rm (Nb_3Sn)}=2.2\lambda_0^{\rm (Nb)}$,  
(c) $\lambda_0^{\rm (NbN)}=4.4\lambda_0^{\rm (Nb)}$, 
and (d) $\lambda_0^{\rm (NbTiN)}=5.3\lambda_0^{\rm (Nb)}$. 
   }\label{fig5}
\end{figure}

%%%%%%%%%%%%%%%%%%
\subsection{Superheating field of a layered superconductor} \label{sec_layered_Hsh}
%%%%%%%%%%%%%%%%%%

Next, we evaluate $H_{sh}$ of the layered structure. 
Since the current density in the layered structure does not necessarily take its maximum value at $x=0$ as seen in Fig.~\ref{fig4}, 
the $H_{sh}$ formula given by Eq.~(\ref{Hsh_formula}) is not applicable. 
Instead, $H_{sh}$ is given by the surface magnetic-field which induces $j_s=j_d^{({\rm S})}$ at $x=0$ or $j_s=j_d^{({\rm \Sigma})}$ at $x=d$. 
Shown as the solid curves in Fig.~\ref{fig5} are $H_{sh}$ as functions of $d$ calculated from the coupled Maxwell-Usadel equations for various materials combinations. 
We find $H_{sh}$ increases with $d$, 
takes its maximum value at $d=d_m \sim \lambda_0^{({\rm S})}$, and decreases with $d$ at $d>d_m$.

Let us interpret these results by using an approximate formula.  
Neglecting the nonlinear Meissner effect and solving the London equation, 
we obtain the well-known formula~\cite{2014_Kubo, 2015_Posen, 2015_Gurevich, 2017_Kubo_SUST}: 
\begin{eqnarray}
&& H_{sh}(d) = {\rm min} \Bigl[ c_1(d) H_{sh}^{\rm (S)} , c_2(d) H_{sh}^{\rm (\Sigma)} \Bigr] , \label{Hsh_London}\\
&& c_1(d) = \frac{\cosh [d/\lambda_0^{\rm (S)}] + [\lambda_0^{\rm (\Sigma)}/\lambda_0^{\rm (S)} ]  \sinh [d/\lambda_0^{\rm (S)}] }{\sinh [d/\lambda_0^{\rm (S)}] + [\lambda_0^{\rm (\Sigma)}/\lambda_0^{\rm (S)} ]  \cosh [d/\lambda_0^{\rm (S)}]} , \label{Hsh_London_1} \\
&& c_2(d) = \cosh [d/\lambda_0^{\rm (S)}] + [\lambda_0^{\rm (\Sigma)}/\lambda_0^{\rm (S)} ]  \sinh [d/\lambda_0^{\rm (S)}]    . \label{Hsh_London_2}
\end{eqnarray}
Here, $H_{sh}^{\rm (S)}$ and $H_{sh}^{\rm (\Sigma)}$ are the superheating field of a semi-infinite superconductor made from the S material and the ${\rm \Sigma}$ material, respectively. 
We use the values obtained from the microscopic theory in the diffusive limit: 
$H_{sh}^{\rm (\Sigma)} =0.795 H_{c0}^{\rm (\Sigma)}$ and $H_{sh}^{\rm (S)} = 0.795 \, r_H  H_{c0}^{\rm (\Sigma)}$ [see Eq.~(\ref{Hsh_zero})]. 
The penetration depth of the S layer in the zero-current state is given by $\lambda_0^{\rm (S)} = \lambda_0^{\rm (\Sigma)} /\sqrt{r_{\Delta} r_{\sigma}}$. 
Shown as the dashed gray curves in Fig.~\ref{fig5} are $H_{sh}(d)$ calculated from Eq.~(\ref{Hsh_London}). 
The existence of the optimum thickness $d_m$ can be understood as follows~\cite{2014_Kubo, 2015_Posen, 2015_Gurevich, 2017_Kubo_SUST}. 
Suppose $\lambda_0^{\rm (S)} > \lambda_0^{\rm (\Sigma)}$. 
Then, the counterflow induced by ${\rm \Sigma}$ decreases the current density at the surface of S by a factor of $1/c_1$, 
and the maximum field that the S layer can withstand increases to $H_0=c_1 H_{sh}^{\rm (S)}$. 
This enhancement is pronounced as $d$ decreases. 
On the other hand, the S layer attenuate the magnetic field down to $H_i = H_0/c_2$ at the S-${\rm \Sigma}$ interface, 
so the maximum field that ${\rm \Sigma}$ can withstand is given by $H_0=c_2 H_{sh}^{\rm (\Sigma)}$, 
which increases with $d$. 
The interplay between the reduction of the surface current and that of the shielding efficiency results in the existence of the optimum thickness $d_m$, 
at which the screening current densities in S and ${\rm \Sigma}$ simultaneously reach $j_d^{({\rm S})}$ and $j_d^{({\rm \Sigma})}$, respectively.

The disagreements between the full calculations (solid) and the approximate formula (dashed) result from the nonlinear Meissner effect. 
The strong current density $\sim j_d$ increases the penetration depth from $\lambda_0$ to $\lambda(s, 0)|_{s\simeq s_d}$, 
so that a larger $d$ becomes necessary to protect ${\rm \Sigma}$ than expected from the London theory. 
As a result, the maximum in $H_{sh}(d)$ obtained from the full calculation is located at thicker regions.

Note $H_{sh}$ can decrease when S has a shorter penetration depth than ${\rm \Sigma}$. 
In this case, the current density in the S layer is enhanced as mentioned in Sec.~\ref{sec_layered_distribution} and can reach the depairing current density at rather small $H_0$. 
For instance, $(r_{\Delta}, r_{H}, r_{\sigma})=(1,1,4)$, which yields $\lambda_0^{\rm (S)}=\lambda_0^{\rm (\Sigma)}/\sqrt{r_{\Delta}r_{\sigma}}=0.5\lambda_0^{\rm (\Sigma)} < \lambda_0^{\rm (\Sigma)}$, 
results in $H_{sh}=0.46 H_{c0}^{\rm (\Sigma)}$ for $d=0.05\lambda_0^{\rm (\Sigma)}$.

%%%%%%%%%%%%%%%%%%%%%%%%%%%%%%%%%%%%
%%%%%%%%%%%%%%%%%%%%%%%%%%%%%%%%%%%%
%%%%%%%%%%%%%%%%%%%%%%%%%%%%%%%%%%%%
%%%%%%%%%%%%%%%%%%%%%%%%%%%%%%%%%%%%
\section{Discussions}
%%%%%%%%%%%%%%%%%%%%%%%%%%%%%%%%%%%%
%%%%%%%%%%%%%%%%%%%%%%%%%%%%%%%%%%%%
%%%%%%%%%%%%%%%%%%%%%%%%%%%%%%%%%%%%
%%%%%%%%%%%%%%%%%%%%%%%%%%%%%%%%%%%%

We have investigated a simple semi-infinite superconductor shown in Fig.~\ref{fig1} (a) and a layered heterostructure shown in Fig.~\ref{fig1} (b) in the diffusive limit. 
The coupled Maxwell-Usadel equations at $T \to 0$ have been self-consistently solved to obtain the spatial distributions of $H(x)$, $j_s(x)$, $\lambda(x)$, and $\Delta(x)$ for both the structures [see Figs.~\ref{fig3} and \ref{fig4}]. 
The distributions of $H$ and $j_s$ obey the London theory for $j_s \ll j_d$, 
while the nonlinear Meissner effect manifests itself for $j_s \sim j_d$, 
where the London theory is no longer valid. 
We have found the superheating field $H_{sh}$ of a semi-infinite superconductor in the diffusive limit is given by $H_{sh} =0.795 H_{c0}$ at $T \to 0$; 
on the other hand, $H_{sh}$ of a layered structure depends on materials combinations and the thickness $d$ of the S layer, 
which can be maximized by tuning $d$ to the optimum thickness [see Fig.~\ref{fig5}].

Our results can be tested by experiments. 
We can expect that the maximum operating field of an SRF cavity made from a bulk dirty-BCS-superconductor is given by its superheating field. 
Taking impurity-doped dirty Nb with $\ell \ll \xi$, for example, 
we have $\mu_0 H_{sh}=0.795 \times 200\,{\rm mT}=160\,{\rm mT}$ at $T\to 0$. 
The maximum operating field can be improved by applying a layered structure onto the inner surface of a cavity. 
For instance, $H_{sh}$ of bulk dirty Nb is pushed up to $\mu_0 H_{sh}=1.04 \times 200\,{\rm mT}=210\,{\rm mT}$ by laminating a thin dirtier-Nb-layer on the surface [see Fig.~\ref{fig5} (a)]. 
Other materials combinations can also improve the field limit, 
e.g., $\mu_0 H_{sh}=2.26\times 200\,{\rm mT}=450\,{\rm mT}$ for the ${\rm Nb_3 Sn}$-Nb structure [see Fig.~\ref{fig5} (b)]. 
These results can be tested using various techniques, e.g., high power rf pulse~\cite{2015_Posen_PRL}, rf characterization of samples~\cite{2015_Welander, 2017_Goudket, 2019_Oikawa, 2019_Oseroff_SRF}, third-harmonic voltage~\cite{2009_Lamura, 2019_Antoine, 2020_Ito, 2019_H_Ito_SRF, 2019_Katayama_SRF}, magnetization measurements for ellipsoid samples~\cite{2016_Tan}, muon-spin-rotation technique~\cite{2018_Junginger, 2019_Keckert}, etc. 
Note that the rf heating of the cavity wall due to quasiparticles~\cite{2019_Kubo_Gurevich, 2020_Kubo_jd, 2014_Gurevich_PRL, 2018_Martinello}, vortices~\cite{2014_Romanenko_flux, 2016_Kubo_flux, 2016_Huang, 2016_Posen, 2016_Checchin, 2018_Liarte_flux, 2019_Miyazaki, 2020_Dhakal, 2020_Pathirana}, topographic defects at the surface~\cite{2008_Iwashita, 2011_Ge, 2013_Yamamoto, 2019_Wenskat, 2020_Pudasaini, 1999_Knobloch, 2015_Kubo_PTEP, 2015_Kubo_PTEP_mag, 2016_Xu}, and grain boundaries~\cite{2017_Ahmad} can limit the achievable field.

The approximate formula [Eq.~(\ref{Hsh_London})], which was derived using the London equation~\cite{2014_Kubo}, 
would be useful to know the optimum thickness $d_{m}$. 
For a clean-limit superconductor at $T=0$, the nonlinear Meissner effect is negligible. 
Hence, simply substituting the clean-limit result $H_{sh}^{(i)}=0.84 H_{c0}^{(i)}$ into Eq.~(\ref{Hsh_London}), 
we obtain the microscopically valid theory~\cite{2015_Gurevich}. 
On the other hand, for a dirty-limit superconductor, the nonlinear Meissner effect is no longer negligible. 
Then, it has been unclear if the results obtained by substituting the dirty-limit result $H_{sh}^{(i)}=0.795 H_{c0}^{(i)}$ into Eq.~(\ref{Hsh_London}) are valid. 
According to our results (see Fig.~\ref{fig5}), the disagreements between the numerical solutions of the coupled Maxwell-Usadel and the approximate formula are $\sim 10\%$, and Eq.~(\ref{Hsh_London}) would still be useful to predict $H_{sh}(d)$ and $d_{m}$ even in the diffusive limit.

%%%%%%%%%%%%%%%
%%%%%%%%%%%%%%%
%acknowledgment
%%%%%%%%%%%%%%%
%%%%%%%%%%%%%%%

\begin{acknowledgments}
I would like to express the deepest appreciation to Alex Gurevich for his hospitality during my visit to Old Dominion University. 
This work was supported by Toray Science Foundation Grant No. 19-6004 and Japan Society for the Promotion of Science (JSPS) KAKENHI Grants No. JP17H04839, JP17KK0100, and JP19H04395. 
\end{acknowledgments}

\appendix

%%%%%%%%%%%%%%%%%%
%%%%%%%%%%%%%%%%%%
\section{Derivations of Eqs.~(\ref{self-consistency_Tzero_2}) and (\ref{superfluid_density_Tzero_2})} \label{a1}
%%%%%%%%%%%%%%%%%%
%%%%%%%%%%%%%%%%%%

We use $\Delta_{0}=\Delta(0,0)$ as a unit of energy. 
The Mastubara Green's function $u=\cot\theta$ for the current carrying state satisfies
\begin{eqnarray}
\biggl( 1-\frac{\zeta}{\sqrt{1+u^2}} \biggr) u = \frac{\omega_n}{\Delta}
\end{eqnarray}
where $\zeta=s/\Delta (s,T)$. 
The self-consistency equation at $T \to 0$ reduces to
\begin{eqnarray}
0 &=& \int_{0}^{\infty} \!\!\! d\omega \biggl( \frac{1}{\Delta \sqrt{1+u^2}}  - \frac{1}{\sqrt{\omega^2 +1}} \biggr) \nonumber \\
&=& \int_{u_0}^{\infty} \!\!\! du \biggl( 1- \frac{\zeta}{(1+u^2)^{3/2}} \biggr) 
\Biggl( \frac{1}{\sqrt{1+u^2}} - \nonumber \\
&& \frac{1}{ \sqrt{\Delta^{-2} + [( 1- \zeta/\sqrt{1+u^2} ) u]^2 } }  \Biggr)  \nonumber \\
&=& -\ln\Delta - \sinh^{-1}u_0 -\frac{\zeta}{2} \biggl( \frac{\pi}{2} -\tan^{-1}u_0 - \frac{u_0}{1+u_0^2}  \biggr)
\label{self-consistency_Tzero_1} \nonumber \\
\end{eqnarray}
Here $u_0(\zeta)$ is defined by $(1-\zeta/\sqrt{1+u_0^2}) u_0 = +0$. 
Eq.~(\ref{self-consistency_Tzero_1}) is the formula of $\Delta$ valid for an arbitrary $s$. 
For $\zeta \le 1$, we have $u_0=0$, and Eq.~(\ref{self-consistency_Tzero_1}) reduces to Eq.~(\ref{self-consistency_Tzero_2}).

The superfluid density $\lambda_0^2/\lambda^2$ can be calculated from Eq.~(\ref{superfluid_density}). 
At $T \to 0$, we find 
\begin{eqnarray}
&&\frac{\lambda_0^2}{\lambda^2(s,0)} 
= \frac{2}{\pi} \int_{0}^{\infty}\!\!\! \frac{d\omega}{1+u^2} \nonumber \\
&&= \frac{2\Delta}{\pi} \int_{u_0}^{\infty} \!\!du \biggl( \frac{1}{1+u^2} - \frac{\zeta}{(1+u^2)^{\frac{3}{2}}} + \frac{\zeta u^2}{ (1+u^2)^{\frac{5}{2}} } \biggr) \nonumber \\
&&=\Delta \biggl[ 1- \frac{2}{\pi}\tan^{-1} u_0 - \frac{4\zeta}{3\pi} \biggl\{ 1 - \frac{u_0 (3+2u_0^2)}{2(1+u_0^2)^{\frac{3}{2}}} \biggr\} \biggr] 
\label{superfluid_density_Tzero_1} , 
\end{eqnarray}
which is the formula of $\lambda$ valid for an arbitrary $s$. 
For $\zeta \le 1$, we have $u_0=0$, 
and Eq.~(\ref{superfluid_density_Tzero_1}) results in Eq.~(\ref{superfluid_density_Tzero_2}).

See also Ref.~\cite{2020_Kubo_Hsh} for generalized formulas which incorporate the effects of a finite Dynes parameter.

\end{document}